\documentclass[12pt]{article}
\usepackage{amssymb,amsmath,epsfig}

\begin{document}
\title{\bf Effects of Electromagnetic Field on the Dynamics of Bianchi
type $VI_0$ Universe with Anisotropic Dark Energy}

\author{M. Sharif \thanks{msharif@math.pu.edu.pk} and M. Zubair
 \thanks{mzubairkk@gmail.com}\\\\
Department of Mathematics, University of the Punjab,\\
Quaid-e-Azam Campus, Lahore-54590, Pakistan.}

\date{}

\maketitle
\begin{abstract}
Spatially homogeneous and anisotropic Bianchi type $VI_0$
cosmological models with cosmological constant are investigated in
the presence of anisotropic dark energy. We examine the effects of
electromagnetic field on the dynamics of the universe and
anisotropic behavior of dark energy. The law of variation of the
mean Hubble parameter is used to find exact solutions of the
Einstein field equations. We find that electromagnetic field
promotes anisotropic behavior of dark energy which becomes isotropic
for future evolution. It is concluded that the isotropic behavior of
the universe model is seen even in the presence of electromagnetic
field and anisotropic fluid.
\end{abstract}


\section{Introduction}

The most remarkable advancement in cosmology is its observational
evidence which says that our universe is in an accelerating
expansion phase. Supernova $I_a$ data \cite{1,2} gave the first
indication of the accelerated expansion of the universe. This was
confirmed by the observations of anisotropies in the cosmic
microwave background (CMB) radiation as seen in the data from
satellite such as WMAP \cite{3} and large scale structure \cite{4}.
Today's one of the major concerns of cosmology is the dark energy.
Recent cosmological observations \cite{1,2,3,4} suggest that our
universe is (approximately) spatially flat and its cosmic inflation
is due to the matter field (dark energy) having negative pressure
(violating energy conditions). The composition of universe density
is the following: 74\% dark energy (DE), 22\% dark matter and 4\%
ordinary matter \cite{5}. Though there is a compelling evidence that
expansion of the universe is accelerating, yet the nature of dark
energy has been under consideration since the last decade
\cite{6}-\cite{10}. Several models have been proposed for this
purpose, e.g., Chaplygin gas, phantoms, quintessence, cosmological
constant and dark energy in brane worlds. However, none of these
models can be regarded as being entirely convincing so far.

The cosmological constant, $\Lambda$ is the most obvious theoretical
candidate of DE which has the equation of state (EoS) $\omega = -1$.
Astronomical observations indicate that the cosmological constant is
many orders of magnitude smaller than estimated in modern theories
of elementary particles \cite{12}. Stabell and Refsdal \cite{14}
discussed the evolution of Friedmann-Lema$\hat{i}$tre Robertson and
Walker (FLRW) dust models in the presence of positive cosmological
constant. These results are given in a more generalized form
\cite{15,16} using the general EoS. Wald \cite{17} examined the late
time behavior of expanding homogeneous cosmological models
satisfying the Einstein field equations (EFEs) with a positive
cosmological constant. He found that all the Bianchi type models
except $IX$ showed isotropic behavior. These models exponentially
evolve towards the de Sitter universe with a scale factor
$(3/{\Lambda})^{1/2}$. Goliath and Ellis \cite{18} used the
dynamical system methods to observe the spatially homogeneous
cosmological models with a cosmological constant. The inclusion of
cosmological constant provides an effective mean of isotropizing
homogeneous universes \cite{17,18}.

The presence of magnetic fields in galatic and intergalatic spaces
is evident from recent observations \cite{19}. The large scale
magnetic fields can be detected by observing their effects on the
CMB radiation. These fields would enhance anisotropies in the CMB,
since the expansion rate will be different depending on the
directions of the field lines \cite{20,21}. Matravers and Tsagas
\cite{24} found that interaction of the cosmological magnetic
field with the spacetime geometry could affect the expansion of
the universe. If the curvature is strong, then even the weak
magnetic field will effect the evolution of the universe. The
magneto-curvature coupling tends to accelerate the positively
curved regions while it decelerates the negatively curved regions
\cite{24,25}.

Jacobs \cite{22} studied the spatially homogeneous and anisotropic
Bianchi type $I$ cosmological model with expansion and shear but
without rotation. He discussed anisotropy in the temperature of
CMB and expansion both with and without magnetic field. It was
concluded that the primordial magnetic field produced large
expansion anisotropies during the radiation-dominated phase but it
had negligible effect during the dust-dominated phase. Dunn and
Tupper \cite{23} discussed properties of Bianchi type $VI_0$
models with perfect fluid and magnetic field. Roy et al. \cite{26}
explored the effects of cosmological constant in Bianchi type $I$
and $VI_0$ models with perfect fluid and homogeneous magnetic
field in the axial direction. They found that model expanded for
negative values of the cosmological constant while it contracted
for positive $\Lambda$. In a recent paper \cite{26a}, Sharif and
Shamir explored the vacuum solution of Bianchi $I$ and $V$ models
in $f(R)$ gravity.

Rodrigues \cite{26b} proposed Bianchi $I$ with a non-dynamical DE
component which yields anisotropic vacuum pressure in two ways: (i)
by considering the anisotropic vacuum consistent with
energy-momentum conservation; (ii) by implementing a Poisson
structure deformations between canonical momenta such that
re-scaling of scale factors is not violated. Koivisto and Mota
\cite{26c} have investigated a cosmological model containing the DE
fluid with non-dynamical anisotropic EoS and interacts with perfect
fluid. They suggested that if the DE EoS is anisotropic, the
expansion rate of the universe becomes direction dependent at late
times and cosmological models with anisotropic EoS can explain some
of the observed anomalies in CMB.

Recently, Akarsu and Kilinc \cite{27} investigated anisotropic
Bianchi type $I$ models in the presence of perfect fluid and
minimally interacting DE with anisotropic EoS parameter. They
found that anisotropy of the DE did not always promote anisotropy
of the expansion. The anisotropic fluid may support isotropization
of the expansion for relatively earlier times in the universe. The
same authors \cite{28} have worked on the Bianchi type $III$ model
in the presence of single imperfect fluid with dynamical
anisotropic EoS parameter and dynamical energy density. They
observed that anisotropy of the expansion vanished and hence the
universe approached isotropy for late times of the universe in
accelerating models.

It would be worthwhile to see what happens if we consider the
electromagnetic field with anisotropic DE. The main purpose of this
work is to look at the effects of electromagnetic field on the
dynamics of the universe in the presence of anisotropic DE for
Bianchi type $VI_0$. The layout of the paper is as follows. In
section \textbf{2} we describe spatially homogeneous and anisotropic
Bianchi type $VI_0$ spacetime and formulate the EFEs in the presence
of anisotropic fluid and magnetic field. Section \textbf{3} presents
a special law of variation for the mean Hubble parameter which
yields constant deceleration parameter. This law generates two types
of solutions: power law and exponential expansion. In section
\textbf{4}, a hypothetical form of fluid is obtained by making an
assumption on anisotropy of the fluid. We obtain exact solutions of
the EFEs and discuss physical behavior of anisotropic DE and
universe model. Finally, section \textbf{5} concludes the results.

\section{Model and the Field Equations}

The spatially homogeneous and anisotropic Bianchi type $VI_0$ model
is described by the line element
\begin{equation}\label{1}
ds^{2}=dt^2-A^2(t)dx^2-e^{2mx}B^2(t)dy^2-e^{-2mx}C^2(t)dz^2,
\end{equation}
where scale factors $A,~B$ and $C$ are functions of cosmic time $t$
only, $m\neq 0$ is a constant. The energy-momentum tensor for the
electromagnetic field is given as \cite{29}
\begin{equation}\label{2}
T_{\mu}^{\nu
(em)}=\bar{\mu}[|h|^2(u_{\mu}u^{\nu}-1/2\delta_{\mu}^{\nu})-h_{\mu}h^{\nu}],
\end{equation}
where $u^{\nu}$ is the four-velocity vector satisfying
\begin{equation}\label{3}
g_{\mu\nu}u^{\mu}u^{\nu}=1.
\end{equation}
$\overline{\mu}$ is the magnetic permeability and $h_{\mu}$ is the
four-magnetic flux given by
\begin{equation}\label{4}
h_{\mu}=\frac{\sqrt{-g}}{2\overline{\mu}}
\varepsilon_{{\mu}{\nu}{\alpha}{\beta}}F^{{\alpha}{\beta}}u^{\nu},\quad
(\mu,\nu,\alpha,\beta=0,1,2,3)
\end{equation}
where $\varepsilon_{{\mu}{\nu}{\alpha}{\beta}}$ is the Levi-Civita
tensor, $F^{{\alpha}{\beta}}$ is the electromagnetic field tensor
and $|h|^ 2=h_{\nu}h^{\nu}$. We assume that magnetic field is due to
an electric current produced along $x$-axis and thus it is in
$yz$-plane. In co-moving coordinates $u^{\nu}=(1,0,0,0)$ and hence
Eq.(4) gives $h_1\neq0,~h_0=h_2=h_3=0$. Using these values in
Eq.(\ref{4}), it follows that $F_{12}=F_{13}=0,~F_{23}\neq0.$

The electric and magnetic field in terms of field tensor are defined
as \cite{30}
\begin{equation}
\label{6} E_{\mu}=F_{{\mu}{\nu}}u^{\nu},\quad B_{\mu}=\frac{1}{2}
\varepsilon_{{\mu}{\nu}{\alpha}}F^{{\nu}{\alpha}}.
\end{equation}
According to Ohm's law, we have
\begin{equation}\label{7}
h_{{\mu}{\nu}}J^{\nu}={\sigma}F_{{\mu}{\nu}}u^{\nu},
\end{equation}
where $h_{{\mu}{\nu}}=g_{{\mu}{\nu}}+u_{\mu}u_{\nu}$ is the
projection tensor orthogonal to $u^{\mu}$,~${\sigma}$ is the
conductivity and $J^{\mu}$ is the four current density. In the
magnetohydrodynamic limit, conductivity takes infinitely large value
while current remains finite so that $E_{\mu}\rightarrow0$
\cite{30}. Consequently, Eq.(\ref{6}) leads to $
F_{01}=F_{02}=F_{03}=0.$ Thus the only non-vanishing component of
electromagnetic field tensor $F_{{\mu}{\nu}}$ is $F_{23}$. The
Maxwell's equations
\begin{equation}\label{8}
F_{{\mu}{\nu};{\alpha}} + F_{{\nu}{\alpha};{\mu}} +
F_{{\alpha}{\mu};{\nu}}=0, \quad F^{{\mu}{\nu}}_{{;} {\alpha}}=0
\end{equation}
are satisfied by
\begin{equation}\label{9}
F_{23}=K=constant.
\end{equation}
It follows from Eq.(\ref{4}) that
\begin{equation}\label{10}
h_1=\frac{AK}{\overline{\mu}BC}, \quad
|h|^2=\frac{K^2}{{\overline{\mu}}^2B^2C^2}.
\end{equation}
Using this equation in Eq.(\ref{2}), we obtain
\begin{equation*}
T_{0}^{{0}{(em)}}=\frac{K^2}{2\overline{\mu}B^2C^2}
=T_{1}^{{1}{(em)}}=-T_{2}^{{2}{(em)}}=-T_{3}^{{3}{(em)}}.
\end{equation*}
Thus we have
\begin{equation}\label{11}
T_{\mu}^{{\nu}{(em)}}=diag[\frac{K^2}{2\overline{\mu}B^2C^2},
\frac{K^2}{2\overline{\mu}B^2C^2},
-\frac{K^2}{2\overline{\mu}B^2C^2},
-\frac{K^2}{2\overline{\mu}B^2C^2}].
\end{equation}

The energy-momentum tensor for anisotropic DE fluid is taken in the
following form \cite{27,28}
\begin{equation}\label{11a}
T_{\mu}^{\nu}=diag[\rho,-p_{x},-p_{y},-p_{z}]
\end{equation}
This model of the DE is characterized by the EoS, $p=\omega{\rho}$,
where $\omega$ is not necessarily constant \cite{31a}. From
Eq.(\ref{11a}), we have
\begin{equation}\label{12}
T_{\mu}^{\nu}=diag[1,-({\omega}+{\delta}),-{\omega},-({\omega}+{\gamma})]\rho,
\end{equation}
where $\rho$ is the energy density of the fluid; $p_x$, $p_y$ and
$p_z$ are pressures and $\omega_x,~\omega_y,~\omega_z$ are
directional EoS parameters on $x,~y$ and $z$ axes respectively.
The deviation from isotropy is obtained by setting
\begin{equation*}
{\omega}_x={\omega}+{\delta}, \quad {\omega}_y={\omega}, \quad
{\omega}_{z}={\omega}+{\gamma},
\end{equation*}
where $\omega$ is the deviation free EoS parameter and $\delta$ and
$\gamma$ are the deviations from $\omega$ on $x$ and $z$ axes
respectively. The EFEs with cosmological constant are given by
\begin{equation}\label{13}
G_{{\mu}{\nu}}=R_{{\mu}{\nu}}-\frac{1}{2}Rg_{{\mu}{\nu}}-{\Lambda}g_{{\mu}{\nu}}
=8\pi(T_{{\mu}{\nu}}+T_{{\mu}{\nu}}^{(em)}),
\end{equation}
where $R_{{\mu}{\nu}}$ is the Ricci tensor, R is the Ricci scalar,
$T_{{\mu}{\nu}}$ is the energy-momentum tensor for anisotropic fluid
and $T_{{\mu}{\nu}}^{(em)}$ is the energy-momentum tensor for the
electromagnetic field.

For Bianchi type $VI_{0}$ spacetime, the EFEs become
\begin{eqnarray}\label{14}
\frac{\dot{A}\dot{B}}{AB}+ \frac{\dot{A}\dot{C}}{AC}
+\frac{\dot{B}\dot{C}}{BC}-\frac{m^2}{A^2}=8{\pi}{\rho}+\frac{4
{\pi}K^2}{\overline{\mu}B^2C^2}+{\Lambda},\\\label{15}
\frac{\ddot{B}}{B}+ \frac{\ddot{C}}{C}
+\frac{\dot{B}\dot{C}}{BC}+\frac{m^2}{A^2}=-8{\pi}({\omega}+{\delta}){\rho}+\frac{4
{\pi}K^2}{\overline{\mu}B^2C^2}+{\Lambda}, \\\label{16}
\frac{\ddot{A}}{A}+ \frac{\ddot{C}}{C}
+\frac{\dot{A}\dot{C}}{AC}-\frac{m^2}{A^2}=-8{\pi}{\omega}{\rho}-\frac{4
{\pi}K^2}{\overline{\mu}B^2C^2}+{\Lambda},\\\label{17}
\frac{\ddot{A}}{A}+ \frac{\ddot{B}}{B}
+\frac{\dot{A}\dot{B}}{AB}-\frac{m^2}{A^2}=-8{\pi}({\omega}+{\gamma}){\rho}-\frac{4
{\pi}K^2}{\overline{\mu}B^2C^2}+{\Lambda}, \\\label{18}
m(\frac{\dot{B}}{B}-\frac{\dot{C}}{C})=0,
\end{eqnarray}
where dot denotes derivative with respect to time $t$. Equation
(\ref{18}) yields
\begin{equation}\label{19}
B=c_1C,
\end{equation}
where $c_1$ is a constant of integration. Subtracting  Eq.(\ref{17})
from (\ref{16}) and using (\ref{19}), we obtain ${\gamma}=0$ which
shows that directional EoS parameters $\omega_y,~\omega_z$ and the
pressures $p_y,~p_z$ become equal. Using Eq.(\ref{19}) and
${\gamma}=0$, the EFEs (\ref{14})-(\ref{17}) reduce to the following
set of equations
\begin{eqnarray}\label{21}
2\frac{\dot{A}\dot{B}}{AB}+ \frac{\dot{B}^2}{B^2}
-\frac{m^2}{A^2}&=&8{\pi}{\rho}+\frac{4
{\pi}K^2}{\overline{\mu}k_1^2B^4}+{\Lambda},\\\label{22}
2\frac{\ddot{B}}{B}+ \frac{\dot{B}^2}{B^2}
+\frac{m^2}{A^2}&=&-8{\pi}({\omega}+{\delta}){\rho}+\frac{4
{\pi}K^2}{\overline{\mu}k_1^2B^4}+{\Lambda}, \\\label{23}
\frac{\ddot{A}}{A}+ \frac{\ddot{B}}{B}
+\frac{\dot{A}\dot{B}}{AB}-\frac{m^2}{A^2}&=&-8{\pi}{\omega}{\rho}-\frac{4
{\pi}K^2}{\overline{\mu}k_1^2B^4}+{\Lambda}.
\end{eqnarray}

\section{Some Physical and Geometrical Parameters}

Here we discuss some physical and geometrical quantities for the
Bianchi type $VI_0$ model which are important in cosmological
observations. The average scale factor is given by
\begin{equation}\label{24}
a=(k_1AB^2)^{\frac{1}{3}}
\end{equation}
while the volume is defined as
\begin{equation}\label{25}
V=a^3=k_1AB^2.
\end{equation}
The mean Hubble parameter $H$ and the directional Hubble parameters
$H_i(i=1,2,3)$ in $x,~y$ and $z$ directions are
\begin{eqnarray}\label{26}
H=\frac{1}{3}({\ln}V\dot{)}={\ln}\dot{a}
=\frac{1}{3}(\frac{\dot{A}}{A}+2\frac{\dot{B}}{B}),\quad
H_x=\frac{\dot{A}}{A}, \quad H_y=H_z=\frac{\dot{B}}{B}.
\end{eqnarray}
The physical parameters such as scalar expansion $\Theta$, shear
scalar $\sigma^2$ and anisotropy of the expansion $\Delta$ are given
as follows
\begin{eqnarray}\label{28}
\Theta&=&u_{; a}^a, \\\label{29}
\sigma^2&=&\frac{1}{2}\sigma_{ab}\sigma^{ab}, \\\label{30}
\Delta&=&\frac{1}{3}\sum_{i=1}^3(\frac{H_i-H}{H})^2.
\end{eqnarray}
The anisotropy of expansion shows isotropic behavior for $\Delta=0$.

It is mentioned here that any universe model becomes isotropic for
the diagonal energy-momentum tensor when
$t\rightarrow{+\infty},~\Delta\rightarrow0,~ V\rightarrow{+\infty}$
and $T^{00}>0~(\rho>0)$ \cite{28,31}. The law of variation of mean
Hubble parameter is given as
\begin{equation}\label{31}
H=la^{-n}=l(k_1AB^2)^{-n/3},
\end{equation}
where $l>0$ and $n\geqslant0$. This law was initially proposed by
Berman \cite{32} for spatially homogeneous and isotropic RW
spacetime which yields constant value of the deceleration
parameter. In recent papers \cite{27,33}, a similar law is
proposed for the homogeneous and anisotropic Bianchi models to
generate exact solutions.

The volumetric deceleration parameter $q$ is the measure of rate at
which expansion of the universe slows down due to self-gravitation.
It is defined as
\begin{equation}\label{32}
q=-\frac{a\ddot{a}}{\dot{a}^2}.
\end{equation}
Using Eqs.(\ref{26}) and (\ref{31}), we get
\begin{equation}\label{33}
\dot{a}=la^{-n+1}, \quad \ddot{a}=-l^2(n-1)a^{-2n+1}
\end{equation}
which gives constant values of the deceleration parameter as follows
\begin{eqnarray*}
q&=&n-1 \quad for \quad n\neq0,\\
q&=&-1  \quad for \quad n=0.
\end{eqnarray*}
Recent observations show that expansion rate of the universe is
accelerating which may be due to the presence of DE and $q<0$. Thus
the sign of $q$ indicates whether the cosmological model inflates or
not. For $q>0$ (i.e. $n>1$), the model represents decelerating
universe where as the negative sign ${-1}{\leqslant}q{<}0$ for
$0{\leqslant}n{<}1$ indicates inflation and $q=0$ for $n=1$
corresponds to expansion with constant velocity. Equations
(\ref{26}) and (\ref{31}) yield two different volumetric expansion
laws
\begin{eqnarray}\label{34}
V&=&k_2e^{3lt},\quad n=0,\\\label{35}
V&=&(nlt+c)^{3/n},\quad n\neq0
\end{eqnarray}
which are used to find exact solutions of the EFEs. In fact these
represent two different models of the universe.

\section{Solution of the Field Equations}

We can find the most general form of anisotropy parameter for the
expansion of Bianchi type $VI_0$ in the presence of anisotropic
fluid and electromagnetic field using Eq.(\ref{26}). The anisotropy
parameter of expansion can be written as
\begin{equation}\label{36}
\Delta=\frac{2}{9H^2}(H_x-H_y)^2,
\end{equation}
where $H_x-H_y$ is the difference between the expansion rates on $x$
and $y$ axes which can be found using the field equations.

Subtracting  Eq.(\ref{23}) from (\ref{22}) and after some
manipulation, it follows that
\begin{equation}\label{37}
\frac{\dot{A}}{A}-\frac{\dot{B}}{B}=\frac{d}{V}
+\frac{1}{V}\int{(8\pi{\delta}{\rho}+\frac{2m^2}{A^2}
-\frac{8{\pi}K^2}{k_{1}^{2}\overline{\mu}B^4})Vdt},
\end{equation}
where $d$ is another constant of integration. Now using this
equation in Eq.(\ref{36}), the anisotropy parameter takes the form
\begin{equation}\label{38}
\Delta=\frac{2}{9H^2}[d+\int{(8\pi{\delta}{\rho}+\frac{2m^2}{A^2}
-\frac{8{\pi}K^2}{k_{1}^{2}\overline{\mu}B^4})Vdt}]^2V^{-2}.
\end{equation}
The anisotropy parameter in the presence of isotropic fluid can be
obtained by choosing $\delta=0$ which yields
\begin{equation}\label{39}
\Delta=\frac{2}{9H^2}[d+\int{(\frac{2m^2}{A^2}
-\frac{8{\pi}K^2}{k_{1}^{2}\overline{\mu}B^4})Vdt}]^2V^{-2}.
\end{equation}
We take the value of $\delta$ so that the integrand in the above
equation vanishes
\begin{equation}\label{40}
\delta=-\frac{m^2}{4{\pi}{\rho}A^2}+\frac{K^2}{k_{1}^2\overline{\mu}{\rho}{B^4}}.
\end{equation}
The corresponding energy-momentum tensor for anisotropic DE fluid
turns out to be
\begin{equation}\label{41}
T_{\mu}^{\nu}=diag[1,-({\omega}-\frac{m^2}{4{\pi}{\rho}A^2}
+\frac{K^2}{k_{1}^2\overline{\mu}{\rho}{B^4}}),-{\omega},-{\omega}]\rho.
\end{equation}
The anisotropy parameter of the expansion reduces to
\begin{equation}\label{42}
\Delta=\frac{2}{9}\frac{d^2}{H^2}V^{-2}.
\end{equation}
We see that $\Delta$ obtained for the Bianchi type $VI_0$ in the
presence of anisotropic fluid with electromagnetic field is
equivalent to that found for the Bianchi type $III$ in the presence
of anisotropic fluid \cite{28}.

The difference between the directional Hubble parameters becomes
\begin{equation}\label{43}
H_x-H_y=\frac{d}{V}.
\end{equation}
The most general form of the energy density is found by using
Eqs.(\ref{21}) and (\ref{30}) as
\begin{equation}\label{44}
\rho=\frac{1}{8{\pi}}[3H^2(1-\frac{\Delta}{2})-\frac{m^2}{A^2}-
{\Lambda}-\frac{4{\pi}K^2}{\overline{\mu}k_{1}^2B^4}].
\end{equation}
This shows that anisotropy of expansion, cosmological constant and
electromagnetic field reduce the energy density $\rho$ of
anisotropic DE.

\subsection{Model for $n=0~(q=-1$)}

The spatial volume of the universe for this model is given by
\begin{equation}\label{45}
V=k_2e^{3lt}.
\end{equation}
Using this value of $V$ in Eq.(\ref{43}) and then solving the EFEs
(\ref{21})-(\ref{23}), the scale factors become
\begin{eqnarray}\label{46}
A&=&k_3e^{lt-\frac{2}{9}\frac{d}{lk_2}e^{-3lt}},\\\label{47}
B&=&(\frac{k_2}{k_{1}k_{3}})^{\frac{1}{2}}e^{lt+\frac{1}{9}\frac{d}{lk_2}e^{-3lt}},\\\label{48}
C&=&(\frac{k_{1}k_{2}}{k_3})^{\frac{1}{2}}e^{lt+\frac{1}{9}\frac{d}{lk_2}e^{-3lt}},
\end{eqnarray}
where $k_1,~k_2$ and $k_3$ are constants of integration.

The directional and the mean Hubble parameters will become
\begin{equation}\label{50}
H_x=l+\frac{2}{3}\frac{d}{k_2}e^{-3lt}, \quad
H_y=H_z=l-\frac{1}{3}\frac{d}{k_2}e^{-3lt},\quad H=l
\end{equation}
while the anisotropy parameter of the expansion takes the form
\begin{equation}\label{51}
\Delta=\frac{2}{9}\frac{d^2}{l^2k_{2}^2}e^{-6lt}.
\end{equation}
The expansion and shear scalar are found as
\begin{eqnarray}\label{52}
\Theta&=&u_{;a}^a=\frac{\dot{A}}{A}+2\frac{\dot{B}}{B}=3l=3H,\\\label{53}
\sigma^2&=&\frac{1}{2}[(\frac{\dot{A}}{A})^2+2(\frac{\dot{B}}{B})^2]-\frac{1}{6}{\Theta}^2
=\frac{1}{3}\frac{d^2}{k_{2}^2}e^{-6lt}.
\end{eqnarray}
The energy density of the DE is evaluated by using Eq.(\ref{21})
with the scale factors as
\begin{eqnarray}\label{54}
\rho&=&\frac{1}{8\pi}[3l^2-\Lambda-\frac{1}{3}\frac{d^2}{k_{2}^2}e^{-6lt}
-\frac{m^2}{k_{3}^2}e^{-2lt+\frac{4}{9}\frac{d}{lk_2}e^{-3lt}}-
\frac{4{\pi}K^2}{\overline{\mu}}(\frac{k_{3}^2}{k_{2}^2})\nonumber\\
&\times& e^{-4lt-\frac{4}{9}\frac{d}{lk_2}e^{-3lt}}].
\end{eqnarray}
The deviation free part of anisotropic EoS parameter $\omega$ can be
obtained by using Eqs.(\ref{46})-(\ref{48}) and (\ref{54}) in
Eq.(\ref{22})
\begin{eqnarray}\label{55}
\omega&=&\{9k_{2}^2k_{3}^2l^2+d^2k_{3}^2e^{-6lt}-3{\Lambda}k_{2}^2k_{3}^2-3m^2k_{2}^2
e^{-2lt+\frac{4}{9}\frac{d}{lk_{2}}e^{-3lt}}+ \frac{12{\pi}k_{3}^4K^2}{\overline{\mu}}\nonumber\\
&\times& e^{-4lt-\frac{4}{9}\frac{d}{lk_2}e^{-3lt}}\}/
\{d^2k_{3}^2e^{-6lt}+3{\Lambda}k_{2}^2k_{3}^2-9k_{2}^2k_{3}^2l^2+3m^2k_{2}^2\nonumber\\
&\times& e^{-2lt+\frac{4}{9}\frac{d}{lk_{2}}e^{-3lt}}+
\frac{12{\pi}k_{3}^4K^2}{\overline{\mu}}
e^{-4lt-\frac{4}{9}\frac{d}{lk_2}e^{-3lt}}\}.
\end{eqnarray}
Using the scale factors and the energy density in Eq.(\ref{40}), the
deviation in EoS parameter along $x$-axis $\delta$ is given as
\begin{eqnarray}\label{56}
\delta&=&\{6m^2k_{2}^2e^{-2lt+\frac{4}{9}\frac{d}{lk_2}e^{-3lt}}+
\frac{24{\pi}K^2k_{3}^4}{\overline{\mu}}e^{-4lt-\frac{4}{9}\frac{d}{lk_2}e^{-3lt}}\}/
\{d^2k_{3}^2e^{-6lt}+3{\Lambda}k_{2}^2 \nonumber\\
&\times&k_{3}^2-9k_{2}^2k_{3}^2l^2+3m^2k_{2}^2e^{-2lt+\frac{4}{9}\frac{d}{lk_{2}}e^{-3lt}}+
\frac{12{\pi}k_{3}^4K^2}{\overline{\mu}}e^{-4lt-\frac{4}{9}\frac{d}{lk_2}e^{-3lt}}\}.
\end{eqnarray}

\subsection{Some Physical Aspects of the Model}

We find that the directional Hubble parameters are dynamical where
as the mean Hubble parameter is constant. Also, the directional
Hubble parameters become constant at $t=0$ and when
$t\rightarrow{\infty}$. These deviate from the mean Hubble parameter
by some constant factor at $t=0$ but coincide when
$t\rightarrow{\infty}$. Since the constant is positive (negative),
it increases (decreases) expansion on the $x$-axis and it decreases
(increases) expansion on $y$ and $z$ axes. The volume $V$ of the
universe is finite at $t=0$, expands exponentially with the increase
in time $t$ and takes infinitely large value as
$t\rightarrow\infty$. Thus the universe evolves with constant volume
and expands exponentially. The expansion scalar is constant for
$0{\leq}t{\leq}{\infty}$ and hence the model represents uniform
expansion.

It is mentioned here that the scale factors $A(t),~B(t)$ and $C(t)$
are finite at $t=0$ which implies that the model has no initial
singularity whereas these diverge for later times of the universe.
The anisotropy parameter of the expansion and shear scalar are found
to be finite for earlier times of the universe where as these
decrease with time and become zero as $t\rightarrow\infty$. This
shows that anisotropy of the expansion is not supported by the
anisotropic DE and electromagnetic field. The quantities
$\rho,~\omega$ and $\delta$ are dynamical and are finite at $t=0$.
The deviation free EoS parameter of the DE may begin in the phantom
$(\omega<-1)$ or quintessence region $(\omega>-1)$ but
$\omega{\rightarrow}-1$ for later times of the universe. One can
observe that $\rho$ increases when $\omega$ is in phantom region and
attains a constant value as $\omega{\rightarrow}-1$. This is shown
in Figure 1.
\begin{figure}
\centering \epsfig{file=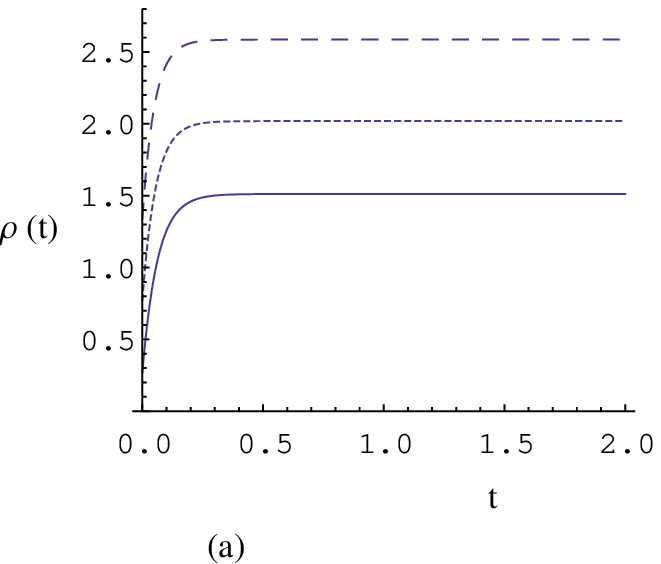,width=.45\linewidth}
\epsfig{file=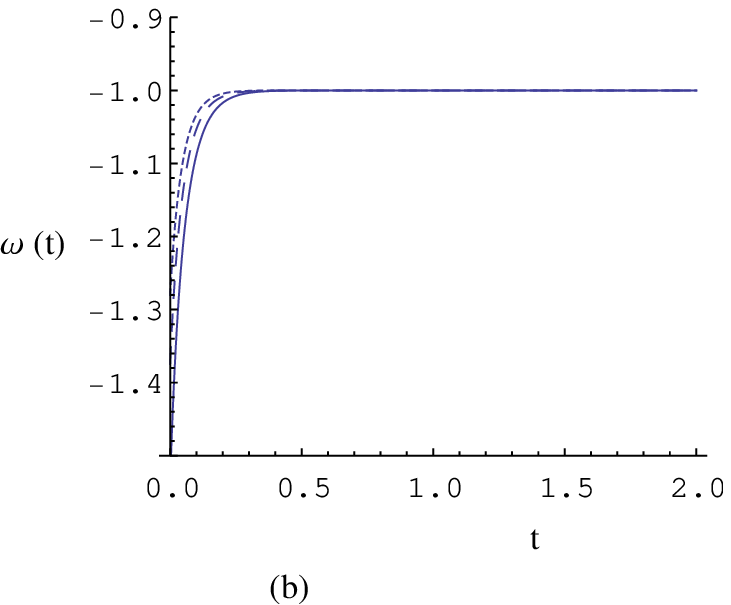,width=.45\linewidth} \caption{Plot of (a)
$\rho(t)$ and (b) $\omega(t)$ verses cosmic time $t$ for $m=1$ and
varying values of $l$ as follows: solid, $l=4$; dotted, $l=4.5$;
dashed, $l=5$.}
\end{figure}

\begin{figure}
\centering \epsfig{file=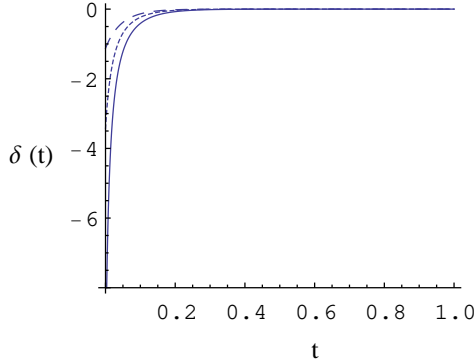,width=.45\linewidth} \caption{Plot
of $\delta$ verses cosmic time $t$ for $m=1$ and varying values of
$l$ as follows: solid, $l=4$; dotted, $l=4.5$; dashed,
$l=5$.}\label{}
\end{figure}

We see from Eq.(\ref{56}) that electromagnetic field favors the
deviation from $\omega$ on $x$-axis, i.e., it contributes to
anisotropic behavior of the fluid. The deviation parameter increases
from negative values towards zero with the increase in time and
tends to zero as $t\rightarrow{\infty}$ shown in Figure 2. Thus the
anisotropic fluid becomes isotropic for the later times of the
universe in the case of exponential volumetric expansion. It follows
from Eq.(\ref{54}) that when $t\rightarrow{\infty},
~{\rho}\rightarrow{3l^2-\Lambda}=3H^2-\Lambda$ which implies that
for $t\rightarrow{\infty},~\rho>0$ only if $H>\sqrt{{\Lambda}/3}$
and hence $\Delta{\rightarrow}0$ and $V\rightarrow{\infty}$. Thus
the model approaches to isotropy for its future evolution. Here
$q=-1,~dH/dt=0$ which gives the largest value of the Hubble
parameter and accelerating rate of expansion.

\subsection{Model for $n\neq0~(q=n-1$)}

The initial time of the universe is found by using Eq.(\ref{35})
\begin{equation}\label{57}
t_*=-c/nl \quad for \quad n\neq0.
\end{equation}
We re-define the cosmic time as
\begin{equation}\label{58}
t'=nlt+c
\end{equation}
such that the initial time turns out to be zero, i.e., $t'=0$. For
this value of cosmic time, we can re-define the Bianchi model in the
form
\begin{equation}\label{59}
ds^{2}=(nl)^{-2}dt'^2-A^2(t')dx^2-e^{2mx}B^2(t')dy^2-e^{-2mx}C^2(t')dz^2.
\end{equation}
The corresponding volume will become
\begin{equation}\label{60}
V=t'^{\frac{3}{n}}.
\end{equation}
Using this value of $V$ in Eq.(\ref{43}) and then solving the EFEs
(\ref{21})-(\ref{23}), the scale factors turn out to be
\begin{eqnarray}\label{61}
A(t')&=&k_3t'^{\frac{1}{n}}e^{\frac{2}{3}\frac{nd}{n-3}t'^{1-\frac{3}{n}}},\\\label{62}
B(t')&=&(\frac{k_2}{k_1k_3})^{\frac{1}{2}}t'^{\frac{1}{n}}
e^{-{\frac{1}{3}\frac{nd}{n-3}t'^{1-\frac{3}{n}}}},\\\label{63}
C(t')&=&(\frac{k_1k_2}{k_3})^{\frac{1}{2}}t'^{\frac{1}{n}}
e^{-{\frac{1}{3}\frac{nd}{n-3}t'^{1-\frac{3}{n}}}}.
\end{eqnarray}

The directional and mean Hubble parameters take the form
\begin{equation}\label{64}
H=(nt')^{-1},\quad
H_x=\frac{1}{nt'}+\frac{2d}{3}t'^{-{\frac{3}{n}}}, \quad
H_y=H_z=\frac{1}{nt'}-\frac{d}{3}t'^{-{\frac{3}{n}}}.
\end{equation}
The corresponding anisotropy parameter of the expansion turns out to
be
\begin{equation}\label{66}
\Delta=\frac{2}{9}n^2d^2t'^{~{2-\frac{6}{n}}}
\end{equation}
while the expansion and shear scalar are
\begin{equation}\label{67}
\Theta=\frac{3}{nt'}=3H,\quad
\sigma^2=\frac{1}{3}d^2t'^{-\frac{6}{n}}.
\end{equation}
The energy density can be found from Eq.(\ref{21}) by using the
scale factors Eq.(\ref{61})-(\ref{63}) as
\begin{eqnarray}\label{69}
{\rho}(t')&=&\frac{1}{8\pi}[3(nt')^{-2}-\Lambda-\frac{1}{3}d^2t'^{-\frac{6}{n}}
-\frac{m^2}{k_{3}^2}t'^{-\frac{2}{n}}e^{-\frac{4}{3}\frac{nd}{n-3}t'^{1-\frac{3}{n}}}
-\frac{4{\pi}K^2}{\overline{\mu}}\nonumber\\
&\times&(\frac{k_{3}^2}{k_{2}^2})t'^{-\frac{4}{n}}
e^{\frac{4}{3}\frac{nd}{n-3}t'^{1-\frac{3}{n}}}].
\end{eqnarray}
Using this equation and the scale factors in Eq.(\ref{22}), we
obtain the deviation free EoS parameter $\omega$
\begin{eqnarray}\label{70}
{\omega(t')}&=&\{(9(nt')^{-2}+d^2t'^{-\frac{6}{n}}-6(nt^2)^{-1}
-3{\Lambda})k_{2}^2k_{3}^2
+3t'^{-\frac{4}{n}}e^{\frac{4}{3}\frac{nd}{n-3}t'^{1-\frac{3}{n}}}\nonumber\\
&\times&(\frac{4{\pi}k_{3}^4K^2}{\overline{\mu}}
-m^2k_{2}^2t'^{\frac{2}{n}}e^{-\frac{8}{3}\frac{nd}{n-3}t'^{1-\frac{3}{n}}})\}/
\{(d^2t'^{-\frac{6}{n}}+3\Lambda-9(nt')^{-2})\nonumber\\
&\times&k_{2}^2k_{3}^2+3t'^{-\frac{4}{n}}e^{\frac{4}{3}\frac{nd}{n-3}
t'^{1-\frac{3}{n}}}(\frac{4{\pi}k_{3}^4K^2}{\overline{\mu}}+
m^2k_{2}^2t'^{\frac{2}{n}}e^{-\frac{8}{3}\frac{nd}{n-3}t'^{1-\frac{3}{n}}})\}.
\end{eqnarray}
Finally, the deviation parameter $\delta$ can be obtained using the
value of ${\rho(t')}$ along with Eqs.(\ref{61})-(\ref{63}) in
(\ref{40})
\begin{eqnarray}\label{71}
{\delta(t')}&=&\{6t'^{-\frac{4}{n}}e^{\frac{4}{3}\frac{nd}{n-3}
t'^{1-\frac{3}{n}}}(\frac{4{\pi}k_{3}^4K^2}{\overline{\mu}}
+m^2k_{2}^2t'^{\frac{2}{n}}e^{-\frac{8}{3}\frac{nd}{n-3}
t'^{1-\frac{3}{n}}})\}/\{(d^2t'^{-\frac{6}{n}}\nonumber\\
&+&3\Lambda-9(nt')^{-2})k_{2}^2k_{3}^2+3t'^{-\frac{4}{n}}
e^{\frac{4}{3}\frac{nd}{n-3}t'^{1-\frac{3}{n}}}
(\frac{4{\pi}k_{3}^4K^2}{\overline{\mu}}+m^2k_{2}^2t'^{\frac{2}{n}}\nonumber\\
&\times&e^{-\frac{8}{3}\frac{nd}{n-3}t'^{1-\frac{3}{n}}})\}.
\end{eqnarray}

\subsection{Some Physical Aspects of the Model}

\begin{figure}
\centering \epsfig{file=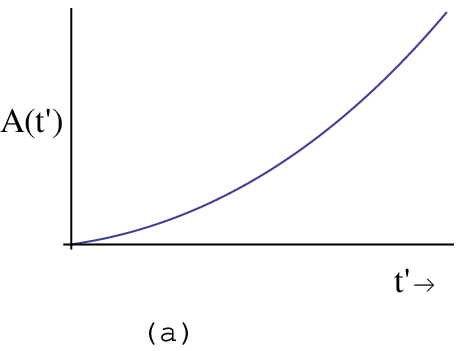} \epsfig{file=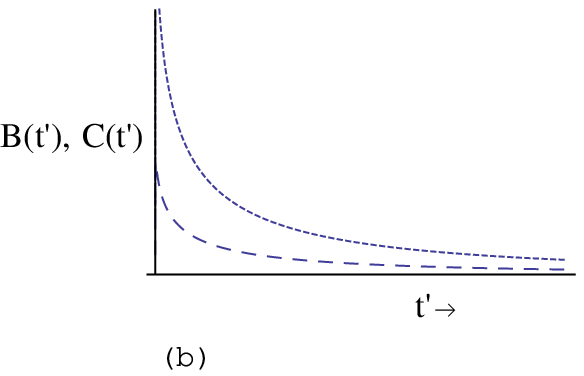}
\caption{Evolution of (a) $A(t')$ and (b) $B(t')$ (dashed line),
$C(t')$ (dotted line) for $n>3,~d>0$ as
$t'\rightarrow{\infty}$.}\label{a}
\end{figure}
The universe model accelerates for $0<n<1$, decelerates for $n>1$
and expands with constant velocity for $n=1$. The mean Hubble
parameter, shear scalar and directional Hubble parameters are
infinite at the initial epoch and tend to zero for later times of
the universe. If $n>3,~d>0$ then $A(t')$ takes infinitely large
value while both $B(t')$ and $C(t')$ vanish as
$t'\rightarrow{\infty}$. This indicates that spacetime exhibits
"pancake" type singularity which is shown in Figure 3. If $n>3,
~d<0$, then $A(t')$ decreases to zero where as both $B(t')$ and
$C(t')$ continue to increase as $t'\rightarrow{\infty}$, leading
to a "cigar" singularity shown in Figure 4. For $n<3$, the scale
factors $A(t'),~B(t')$ and $C(t')$ become infinite as
$t'\rightarrow{\infty}$ given in Figure 5.

\begin{figure}
\centering \epsfig{file=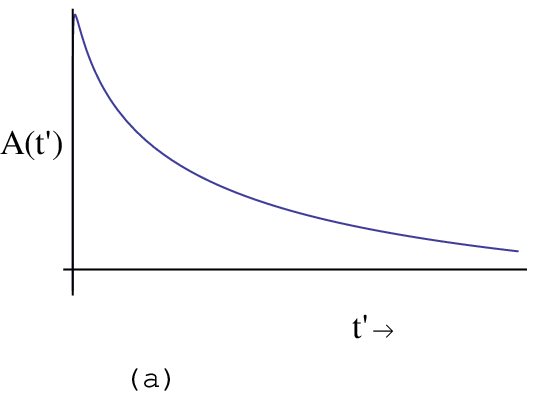} \epsfig{file=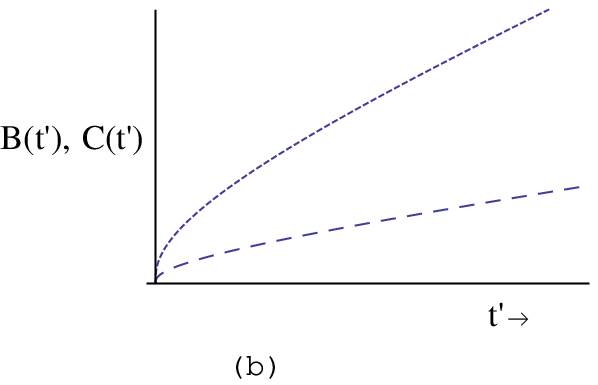}
\caption{Evolution of (a) $A(t')$ and (b) $B(t')$ (dashed line),
$C(t')$ (dotted line) for $n>3$, $d<0$ as $t'\rightarrow{\infty}$.}
\end{figure}

\begin{figure}
\centering \epsfig{file=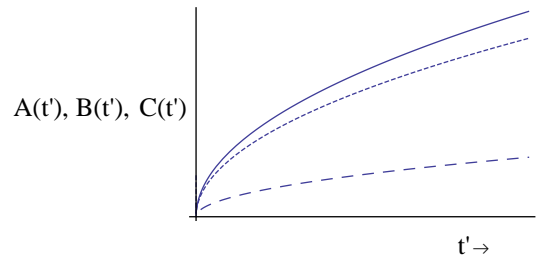} \caption{Evolution of $A(t')$, solid
line; $B(t')$ dashed; $C(t')$ dotted for $n<3$ as
$t'\rightarrow{\infty}$.}
\end{figure}

We note that spatial volume $V$ is zero at $t'=0$ and it takes
infinitely large value as $t'\rightarrow{\infty}$. The expansion
scalar is infinite at $t'=0$ and decreases with the increase in
cosmic time. Thus the universe starts evolving with zero volume at
the initial epoch with an infinite rate of expansion and expansion
rate slows down for the later times of the universe. The dynamics of
anisotropy parameter of the expansion $\Delta$ depends on the value
of $n$. $\Delta$ tends to zero as $t'\rightarrow{\infty}$ and
diverges as $t'\rightarrow0$ for $n<3$ and vice versa for $n>3$
while it remains constant for $n=3$ shown in Figure 6.

Now we determine such values of $n$ which satisfy $\rho>0$ and are
suitable for the evolution of the universe. We consider magnetic
field, anisotropy of expansion and cosmological constant for the
behavior of energy density of the DE. For this purpose, we discuss
the following two cases, i.e., $n<3$ and $n>3$. When $n<3$, we have
$\rho<0$ as $t'\rightarrow0$ and hence this model is not suitable
for representing the relatively earlier times of the universe. When
$t'\rightarrow{\infty}$, it follows that ${\rho}\geqslant0$ with
${\Lambda}\leqslant0$ which is an appropriate model for representing
the later times of the universe. For $n>3$, we obtain $\rho<0$ as
$t'\rightarrow{\infty}$ and it becomes positive as $t'\rightarrow0$.
Thus this model can represent the universe only for the earlier
times of the universe by assigning suitable values to the constants.

\begin{figure}
\centering
\epsfig{file=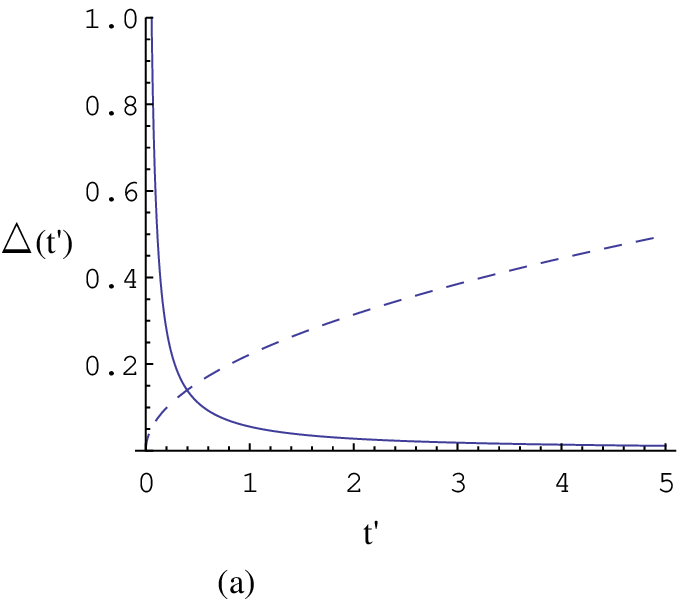}
\epsfig{file=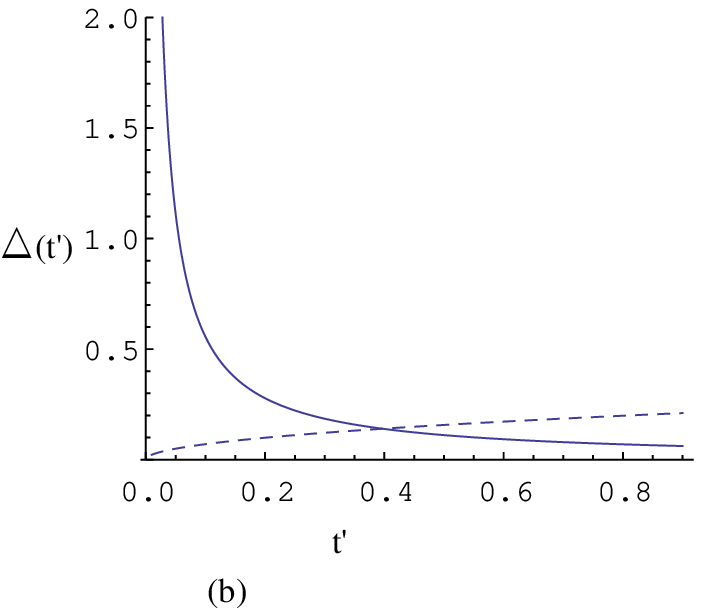} \caption{Plot of $\Delta$ verses cosmic time
$t'$; solid line for $n<3$ and dotted for $n>3$ as (a)
$t'\rightarrow{\infty}$ (b) $t'\rightarrow0$}
\end{figure}

If $n<3$ and $\Lambda{\leqslant0},~\rho{\geqslant}0$ (for later
times of the universe), $\omega$ begins in quintessence region and
then passes into the phantom region. It remains in the phantom
region with the increase in time and becomes $-1$ as
$t'\rightarrow{\infty}$. If $n>3$ with $\rho{\geqslant}0$ (for
earlier times of the universe), $\omega$ begins in quintessence
region, passes into the phantom region for small interval of time
then it passes back into the quintessence region and becomes $-1$ as
$t'\rightarrow{\infty}$.

Thus we may examine the behavior of anisotropy of the fluid and
universe model for $n<3$ with $\Lambda{\leqslant0}$ and
$\rho{\geqslant}0$ as $t'\rightarrow{\infty}$. One can observe that
magnetic field may increase anisotropic behavior of the DE since it
contributes to $\delta$ which decreases with the increase in time
and it tends to zero as $t'\rightarrow{\infty}$. Thus the anisotropy
of the DE vanishes in the presence of magnetic field for later times
of the universe with negative cosmological constant. We note that
anisotropy parameter of the expansion is not supported by anisotropy
of the DE and magnetic field for later times of the universe since
for $n<3,~\Delta{\rightarrow}0$ as $t'\rightarrow{\infty}$. It is
observed that $V{\rightarrow}\infty$ and $\rho{\geqslant}0$ as
$t'\rightarrow{\infty}$ for $n<3$ and hence the model represents
isotropic universe for future evolution.

\section{Summary and Conclusion}

We have obtained two exact solutions of the dynamical equations
for the spatially homogeneous and anisotropic Bianchi type $VI_0$
model with magnetic field, anisotropic DE and cosmological
constant. The dark energy component is dynamical which yield
anisotropic pressure. Assuming the law of variation of the mean
Hubble parameter, the cosmological models are given for $n=0$ and
$n\neq0$. The physical and geometrical properties of the models
are discussed. We have found the explicit form of scale factors
and have explained the nature of singularities.

The model represents uniform expansion for the exponential expansion
while in the case of power law expansion, universe expands with an
infinite rate of expansion which slows down for the later times of
the universe. It is found that electromagnetic field affects
anisotropies in the CMB, in particular, it increases anisotropic
behavior. Our results show that even the fluid is anisotropic which
yields anisotropic EoS parameter with the electromagnetic field, its
anisotropy vanishes for future evolution of the universe in both
cases. The expansion of the universe becomes isotropic due to the
isotropic behavior of the fluid when $t'\rightarrow{\infty}$.

The model with zero deceleration parameter can approach to
isotropy as $t'\rightarrow{\infty}$ with the condition that
$H>\sqrt{\Lambda/3}$, $\Lambda{\geqslant}0$. It is shown that
$\omega$ is in the phantom region which tends to constant value
$-1$ for later times of the universe. Thus the expanding model is
accelerating. Bianchi models usually isotropize for the positive
cosmological constant but we have shown that model for the power
law expansion isotropizes for $n<3,~\Lambda{\leqslant}0$ for later
times of the universe. The above analysis shows that inclusion of
the cosmological constant in homogeneous cosmological model
greatly affects the late time behavior as pointed out by Ellis
\cite{18}. It is interesting to mention here that though $\omega$
begins in quintessence region then passes into the phantom region,
but it remains in the phantom region and tends to $-1$ as
$t'\rightarrow{\infty}$ which can result in accelerating the
expansion. However, due to the presence of negative cosmological
constant, it starts contracting. One can observe that this
behavior may re-collapse the universe and hence the model
represents decelerating expansion of the universe.

Finally, we would like to mention here that the model (the
exponential expansion law) represents expanding universe for both
present and future evolution which fits with the current
observations and hence the standard $\Lambda$CDM model. Also, the
observational data favors the power law $\Lambda$CDM model \cite{1}.
We have found that the model represents an accelerating universe for
the power law expansion. Both the expansion models approach to the
EoS of cosmological constant for future evolution. The anisotropy of
the universe and DE vanish for the period of the accelerated
expansion. Thus the isotropy is observed for the future evolution of
the universe. However, there is still a possibility of DE component
with anisotropic EoS in present epoch of the universe.

\end{document}